# On geodesics of gradient-index optical metrics and the optical-mechanical analogy

David Delphenich
Spring Valley, OH 45370 USA

___

**Abstract:** The geodesic equations for optical media whose refractive indices have a non-vanishing gradient are developed.  It is shown that when those media are optically isotropic, the light paths will be the null geodesics of a spatial metric that is conformally related to the metric on the ambient space.  Various aspects of the optical-mechanical analogy are discussed as they relate to the geodesics of conformally-related metrics.  Some applications of the concepts of gradient-index optics to general relativity are examined, such as effective indices of refraction for gravitational lensing and Gordon's optical metrics for optical media that are in a state of relative motion with respect to an observer.  The latter topic is approached from the standpoint of pre-metric electromagnetism.

**Contents**

Page

**1. Introduction.** – The path of a light ray in an optical medium is a geodesic of the spatial metric that prevails in that medium.  In the most common optical media, those geodesics will be straight lines that typically change directions only at the interface with another medium.  That is simply because the most common optical media have indices of refraction that are constant throughout the medium and then changes only discontinuously at the boundaries of the regions that they occupy.  Hence, the effective metric that one obtains the geodesics from will be only a constant times the usual Euclidian metric that defines the geometry of spaces in which one does not have to deal with relative velocities comparable to the speed of light (except, of course, the light rays themselves!) or strong gravitational fields.

In order to get light rays that are not straight lines, one must deal with inhomogeneous optical media in which the indices of refraction change continuously (or rather, they are continuously differentiable) throughout the medium. In such a case, the effective metric that one considers will be conformal to the Euclidian metric, and the conformal factor will be the square of the index of refraction, at least for isotropic, time-invariant optical media. The equations that define the geodesics for such a metric will then have non-vanishing connection coefficients that include contributions from the gradient of the index.

Gradient-index optical media [**1**] currently exist in the technological sense of the term, although the processes for manufacturing them are rather advanced, so they have come about only in the last half-century. Some of the technological applications of gradient-index media include the fact that one can replace a bundle of homogeneous optical fibers with a single inhomogeneous one, and such an approach has been implemented in order to make endoscopes that are used in medical applications.

However, a naturally-occurring example has been around for quite some time in the form of the Earth's atmosphere. In particular, the index of refraction seems to vary most strongly with the temperature of the atmosphere (or really, with the density). That explains the way that hot surfaces, such as road surfaces, can appear to be reflecting, since the fact that the index of refraction is lower down close to the road surface than it is above it implies that one will get total internal reflection at some critical angle of incidence. Similarly, mirages and *fata morganas* come about when the index of refraction of the air changes in such a way that it has a non-vanishing gradient. The fact that the Earth's atmosphere tends to distort the images of objects outside the atmosphere when they are close to the horizon has been known since the time of the Mesopotamian scientist Al Hazen (also known as Ibn al Haytham, who lived around 1100 A.D.), and that can be explained by the optical inhomogeneity of the atmosphere, as well.

The eyes of animals are also optically inhomogeneous, in general.

One can also find applications of gradient-index optics to the space-time of general relativity. One of the more commonplace applications is due to the fact that strong gravitational fields, such as one might expect around dense stellar objects like neutron stars and black holes, are known to deflect light rays that pass close to them. That experimentally-established prediction of general relativity leads to another experimentally-established phenomenon that is referred to as "gravitational lensing." That phenomenon is the basis for observing the presence of black holes, since although they emit no light of their own, their presence distorts the optics of space in their vicinity in a manner that allows astronomers to infer their existence. There have been various attempts to model gravitational lensing in terms of an effective index of refraction for an optically-inhomogeneous medium [**2-9**]. Indeed, some theoreticians believe that the intense magnetic fields of magnetars, which are a type of neutron star, might even lead to the optical phenomenon that is known as birefringence for polarized light that passes close to them.

Yet another example of a medium that would have an effective index of refraction that comes from the theory of general relativity goes back to the German physicist Walter Gordon (of Klein-Gordon fame) in 1923 [**10**]. He showed that when an optical medium is in a state of relative motion with respect to an observer, that relative velocity will define an effective metric for the medium that is associated with an effective polarization.

That fact relates to one of the subtle aspects of the theory of relativity, namely, that although Einstein's work largely concluded with a theory of strong gravitational fields, it started by examining the propagation of electromagnetic waves in moving media. Indeed, that was the starting point for Max von Laue's approach to the theory of relativity [**11**].

It is also at the heart of the approach to the theory of "pre-metric" electromagnetism [**12, 13**], which is based upon the observation that the only place where a space-time metric gets used in the Maxwell equations is in the context of the electromagnetic constitutive law that associates the response of the medium (i.e., the electric displacement or excitation and the magnetic flux density or excitation) with the fields that are imposed upon it (i.e., the electric and magnetic field strengths). The Lorentzian metric of gravitation then arises as a corollary to the dispersion law for the propagation of electromagnetic waves that is defined by some fairly restrictive special-case conditions.

The structure of this study of geodesics in gradient-index optical media will be to first review some basic ideas in Section **2** that are associated with geodesics in curved space and how Fermat's principle of optics relates to them. The fact that when the optical medium is time-invariant and isotropic, the optical metric that is defined by the index of refraction is conformally-related to the metric of the ambient space in which the medium is embedded will then be discussed in Section **3**. In Section **4**, some of the analogies between light rays and the trajectories of point masses moving under the influence of conservative external forces are reviewed. In Section **5**, some of the applications of gradient-index optics to general relativity, such as gravitational lensing and Gordon's optical metric, will be presented, and the results will be summarized in section **6**.

To some extent, this paper is a survey of existing results from sources that come from specialties whose researchers might not always be aware of each other's existence. However, the novelty in the treatment of those topics is perhaps in its use of modern geometric methods and its focus on the way that conformal transformations pervade many branches of physics. Furthermore, in the treatment of the Gordon optical metric, there is a shift the emphasis from the usual metric theory of gravitation to the pre-metric theory of electromagnetism.

Although a certain familiarity with differential geometry, as it is usually employed in general relativity, will be assumed, every attempt has been made to make this study self-contained in regard to the mathematical topics. However, in the discussion of Gordon's optical metrics, some familiarity with multilinear algebra, and especially exterior algebra, will be necessary.

**2. Geodesics and Fermat's principle.** – Let us first recall the geometric nature of geodesics and how Fermat's principle alters their definition.

*a. Geodesics*. – Generally, in the purely-geometric context one can obtain the equation of geodesics in a curved space by various means, one of which is by defining them to be paths of least distance between finitely-spaced points in the space. That can be formulated as a problem in the calculus of variations that involves finding the extremals of an action functional that takes the form of the total arc-length of the curve:

$$S[x(s)] = \int_{x(s_0)}^{x(s_0)} ds \, . \tag{2.1}$$

The expression $S[x(s)]$ suggests that the action functional associates entire curves $x(s)$ with numbers, not just points in space, and the limits of the integral $x(s_0)$ and $x(s_1)$ are the two fixed points in space between which one seeks the shortest path. The symbol $ds$ refers to the differential element of arc length for the curve $x(s)$, which is generally represented as a homogeneous quadratic form in the differentials $dx^i$ of the coordinates $x^i(s)$ of a typical point along the curve, namely:

$$ds^2 = g_{ij}(x) \, dx^i \, dx^j \, . \tag{2.2}$$

We are allowing the components of the metric $g_{ij}(x)$ to vary in space, not only for the sake of generality, but because it will also become necessary when we introduce an index of refraction with a non-vanishing gradient. However, in the absence of strong gravitational fields, the components $g_{ij}$ are usually those of the Euclidian metric $\delta_{ij}$, which equals 1 when $i = j$ and 0 otherwise.

As it stands, the Lagrangian that would be most directly relevant to the action functional in (2.1) would equal just 1, which tends to disguise the complexity of the problem. In order to get around that, one replaces the arc-length parameter $s$ with an arbitrary parameter $\sigma$ by the following rule:

$$s = s(\sigma), \qquad ds = v(\sigma) \, d\sigma, \qquad v = \frac{ds}{d\sigma} \, . \tag{2.3}$$

Since the change of parameter is assumed to be invertible and differentiable and have a differentiable inverse (i.e., it is a *diffeomorphism*), $v(\sigma)$ cannot vanish for any $\sigma$, and in fact it must remain positive in order to preserve the sense of advance along the curve. $v$ is referred to as the *speed* of the parameterization $\sigma$ with respect to arc-length. In particular, when $\sigma = t$ (viz., time), it will be the speed in the usual sense.

With that substitution, (2.1) will take the form:

$$S[x(\sigma)] = \int_{x(\sigma_0)}^{x(\sigma_1)} v(\sigma) \, d\sigma \, , \tag{2.4}$$

so the Lagrangian for this action functional will now take the more illuminating form:

$$\mathcal{L}(x, \dot{x}) = [g_{ij}(x) \, \dot{x}^i \, \dot{x}^j]^{1/2} \, , \tag{2.5}$$

in which the dots signify differentiation with respect to $\sigma$.

The usual process for finding extremals of $S[x(\sigma)]$ is to look for the curves $x(\sigma)$ for which the first variation of that functional vanishes for all variations $\delta x(\sigma)$ (also called "virtual displacements") of the curve, and usually under the assumption that the variations $\delta x(\sigma)$ must vanish at the endpoints of the curve: $\delta x(\sigma_0) = \delta x(\sigma_1) = 0$. The most concrete way of imagining a variation is to think of it as simply a vector field along the curve that is not tangential to it anywhere. (Tangency would suggest that one was also varying the parameterization itself.)

The ultimate outcome of the aforementioned process is the Euler-Lagrange equations for the extremal curve:

$$0 = \frac{\delta \mathcal{L}}{\delta x^i} \equiv \frac{\partial \mathcal{L}}{\partial x^i} - \frac{d}{d\sigma}\left(\frac{\partial \mathcal{L}}{\partial \dot{x}^i}\right). \tag{2.6}$$

In general, if the space is *n*-dimensional then this will represent a system of *n* second-order ordinary differential equations for the curve $x^i(s)$ whose complete solution will then require $2n$ pieces of independent data. Typically, one obtains them by posing either an initial-value problem, for which the data come from the position coordinates and velocity components at some initial value $\sigma_0$ of the parameter $\sigma$, or by posing a two-point boundary-value problem, for which the data might come from the coordinates of the starting point and the endpoint.

For the Lagrangian that was defined in (2.5), the Euler-Lagrange equations will initially take the form:

$$0 = \frac{1}{2v}\frac{\partial g_{jk}}{\partial x^i}\dot{x}^j\dot{x}^k - \frac{d}{d\sigma}\left(\frac{1}{v}g_{ij}\dot{x}^j\right) = \frac{1}{2v}\frac{\partial g_{jk}}{\partial x^i}\dot{x}^j\dot{x}^k - \frac{d}{d\sigma}\left(\frac{1}{v}\right)g_{ij}\dot{x}^j - \frac{1}{v}\frac{dg_{ij}}{d\sigma}\dot{x}^j - \frac{1}{v}g_{ij}\ddot{x}^j.$$

Since the parameterization was arbitrary, we can now revert back to arc-length parameterization in order to simplify this. Hence, we set $\sigma = s$, so $v = 1$, and the derivative of $v$ will vanish. That makes the last equation now take the form:

$$0 = \tfrac{1}{2}\partial_i g_{jk}\dot{x}^j\dot{x}^k - \frac{dg_{ij}}{ds}\dot{x}^j - g_{ij}\ddot{x}^j = \tfrac{1}{2}\partial_i g_{jk}\dot{x}^j\dot{x}^k - \partial_k g_{ij}\dot{x}^j\dot{x}^k - g_{ij}\ddot{x}^j,$$

in which we have abbreviated the partial derivative $\partial / \partial x^i$ by simply $\partial_i$. The last expression is still not in its final form, because the partial derivative in the middle term needs to be symmetrized with respect to the subscripts *j* and *k* in order to be consistent with the product of the velocities. Ultimately, we will get an equation of the form:

$$0 = -\frac{\delta v}{\delta x^i} = g_{ij}\ddot{x}^j + [i, jk]\dot{x}^j\dot{x}^k, \tag{2.7}$$

into which we have introduced the *Christoffel symbols of the first kind:*

$$[i, jk] \equiv \tfrac{1}{2}[\partial_j g_{ik} + \partial_k g_{ij} - \partial_i g_{jk}]. \tag{2.8}$$

Since the component matrix $g_{ij}$ for the spatial metric is assumed to be invertible, we can essentially factor it out of both terms on the right-hand side of (2.7) and get the contravariant form for the geodesic equations:

$$0 = \frac{d^2 x^i}{ds^2} + \left\{ \begin{matrix} i \\ j\, k \end{matrix} \right\} \frac{dx^j}{ds} \frac{dx^k}{ds}, \tag{2.9}$$

into which we have introduced the *Christoffel symbols of the second kind:*

$$\left\{ \begin{matrix} i \\ j\, k \end{matrix} \right\} \equiv g^{il}\, [l, jk] = \tfrac{1}{2} g^{il}\, (\partial_j\, g_{ik} + \partial_k\, g_{ij} - \partial_i\, g_{jk})\,. \tag{2.10}$$

One sees that these symbols will all vanish for the Euclidian case (or indeed for any metric whose components are constant in space). The geodesic equation will then reduce to simply the vanishing of the acceleration of the curve (with respect to $s$), and the solutions to that geodesic equation will be straight lines.

The Christoffel symbols of either kind define the *Levi-Civita connection* that is defined by the metric $g$. It is not an actual tensor field, but a "1-form with values in the Lie algebra of infinitesimal rotations," and it specifies how an orthonormal frame field along the curve $x$ ($s$) will get rotated under "parallel translation along the curve." If the reader is not familiar with those concepts then they can relax because we shall not always employ them in full generality, although the curious can confer various references on the subject of modern differential geometry and its applications to physics (e.g., Frenkel [**14**]).

We introduce the notation for covariant differentiation of a vector field **X** ($s$) = $X^i$ ($s$) $\partial_i$ ([1]) along a curve parameterized by $s$ using the Levi-Civita connection:

$$\nabla_s X^i = \frac{dX^i}{ds} + \left\{ \begin{matrix} i \\ j\, k \end{matrix} \right\} \frac{dx^j}{ds} X^k, \tag{2.11}$$

and when this vanishes, one says that the vector field **X** ($s$) is *parallel* along the curve $x$ ($s$) or *parallel translated* along it.

Now observe that the geodesic equation (2.9) can then be put into the form:

$$0 = \nabla_s \dot{x}^i, \tag{2.12}$$

which says that for the velocity (or unit tangent) vector field of a geodesic is parallel-translated along it or that its acceleration vanishes.

One can give the geodesic equations a slightly more geometrically-intuitive interpretation by introducing a *Frenet frame field* along the curve $x$ ($s$), at least for the segments of it that are not straight lines (i.e., ones for which the acceleration is collinear with the velocity). We shall again revert to a Euclidian metric on the ambient space, so covariant differentiation will reduce to $d/ds$. We shall then write the scalar product of two vectors **v**, **w** in the form $<$ **v**, **w** $>$.

---

([1]) Some readers might not be familiar with the representation of tangent vectors by directional derivatives, which is commonplace in the modern treatments of differentiable manifolds (see, e.g., [**15**]). However, it does have the advantage that when a 1-form, such as a coordinate differential $dx^i$, is applied to a tangent vector as a linear functional, the differential of $x^i$ will become reciprocal to the directional derivative $\partial_i$ in the direction of the coordinate $x^i$ : $dx^i$ ($\partial_j$) = $\delta^i_j$.

Since the speed of the arc-length parameterization is unity, the tangent vector field **t** ($s$) that is defined by the velocity vector field for the curve, namely:

$$\mathbf{t}(s) \equiv \frac{dx}{ds} = \frac{dx^i}{ds}(s)\partial_i,  \quad (2.13)$$

will be a unit vector field, so:

$$<\mathbf{t}, \mathbf{t}> = 1. \quad (2.14)$$

When one differentiates that with respect to $s$, one will find that the derivative of **t** is orthogonal to **t**:

$$<\mathbf{t}, \frac{d\mathbf{t}}{ds}> = 0. \quad (2.15)$$

The acceleration vector field $d\mathbf{t}/ds$ is not itself a unit vector field, but as long as it is non-vanishing, one can always find a scalar function of arc-length $\kappa(s)$ that allows one to express it as a scalar multiple of a unit vector field **n** ($s$):

$$\frac{d\mathbf{t}}{ds} = \frac{d^2 x}{ds^2} = \kappa \mathbf{n}. \quad (2.16)$$

The scalar function $\kappa(s)$ is called the *curvature* of the curve $x(s)$, and the unit vector field **n** ($s$) is the *normal vector field*.

One can continue on to other mutually-orthogonal unit vector fields that will ultimately define an orthonormal $n$-frame field along the curve, but we shall not use any of them besides **t** ($s$) and **n**($s$). That is because the next vector field that one defines (viz., the *binormal*) will be coupled to the derivative of **n** with respect to $s$, which will then involve third derivatives of $x(s)$ with respect to $s$, and the geodesic equations are only second-order ordinary differential equations. However, when one has obtained a solution to the geodesic equations, one can always continue to differentiate (assuming that the derivatives exist) and then discuss its *torsion*. The plane in each tangent space $T_{x(s)}M$ that is spanned by the vectors **t** and **n** is called the *osculating plane* to the curve $x(s)$ at that point.

One can then put the geodesic equation (2.9) into the form:

$$\kappa n^i = -\begin{Bmatrix} i \\ j\ k \end{Bmatrix} t^j t^k, \quad (2.17)$$

which says that contribution to the covariant derivative of the tangent vector field along a geodesic that comes from the connection is always normal to it. Hence, the geodesic equation also says that the tangential component of the acceleration must always vanish, which would be typical of unforced motion.

If one contracts both sides of (2.17) with the covariant form of the unit normal field $n_i$ ($s$) [i.e., takes the scalar product of both sides with the vector field **n** ($s$)] then the last equation will take the form:

$$\kappa = -n_i \begin{Bmatrix} i \\ j\ k \end{Bmatrix} t^j t^k = -\delta_{il} \begin{Bmatrix} i \\ j\ k \end{Bmatrix} t^j t^k n^l . \tag{2.18}$$

That says that the curvature of the curve $x$ ($s$) can be obtained from a quadratic form in the unit tangent vectors. Once again, when the Christoffel symbols vanish, so will the curvature, and the geodesic will be a straight line.

*b. Fermat's principle* [**16-18**]. – What Fermat noticed about light rays in refractive media was that the curves that they defined were not so much paths of least *distance*, but paths of least *duration*. Since the differential element of time $dt$ along a curve that is parameterized by time $t$ relates to the differential element of arc-length $ds$ by the simple rule:

$$ds = v\ (t)\ dt, \tag{2.19}$$

one sees that the points traced out by paths of least time will coincide with those of the paths of least distance and the paths will differ only in the speeds of their parameterizations.

However, in an optical medium, the speed of a light ray will be the speed of propagation of light waves in that medium at each point, which we shall call $c$, and it in turn defines the *index of refraction* of the medium $n$ by way of:

$$n = \frac{c_0}{c}, \tag{2.20}$$

in which $c_0$ is the speed of light in the classical vacuum, which is the regarded as an upper bound on $c$; hence, indices of refraction will always be greater than or equal to 1. The variables upon which $c$ (and therefore $n$) can depend vary with the medium in question. In particular, they can vary with time, position, and direction of the light ray. Since the tangent vector field for a curve that is parameterized by arc-length is a unit vector field, one can regard **t** ($s$) as a variable that defines the direction in space. The index of refraction can also vary with the polarization state of a light wave, but we shall not elaborate upon that topic here.

If one uses $c$ for $v$ above and rewrites (2.20) in the form $c = c_0 / n$ then the change of parameter from arc-length to time (2.19) can be expressed in the form

$$n\ ds = c_0\ dt . \tag{2.21}$$

When the optical medium is time-invariant and isotropic, so $n = n$ ($x$), one can also think of the left-hand side of this expression as defining a new arc-length $\bar{s}$ for the same curve that relates to the geometric one $s$ by way of:

$$\frac{d\bar{s}}{ds} = n\ (s) . \tag{2.22}$$

The arc-length parameter $\bar{s}$ defines what is commonly called the *optical path length* [18].

Since $c_0$ is a constant, a curve that is extremal for the action functional that expresses the elapsed time along the curve $x(t)$ :

$$T[x(t)] = \int_{x(t_0)}^{x(t_0)} dt \qquad (2.23)$$

will also be extremal for one that has $c_0 \, dt$ as an integrand. Hence, from (2.21), one can express the elapsed-time functional in the form:

$$T[x(s)] = \int_{x(s_0)}^{x(s_0)} n \, ds, \qquad (2.24)$$

which can also be written in terms of optical path length:

$$T[x(\bar{s})] = \int_{x(\bar{s}_0)}^{x(\bar{s}_0)} d\bar{s}, \qquad (2.25)$$

as long as the medium is isotropic. One will then see that the only essential difference between the elapsed-time functional and the arc-length functional is the introduction of a scalar multiple in the form of $n$.

Hence, with the reparameterization of the curve $x(s)$ to $x(\sigma)$ that replaces $ds$ with $v \, d\sigma$, as before, if we confine ourselves to media that are inhomogeneous and anisotropic, but time-invariant, then the new Lagrangian will take the form:

$$\mathcal{L}(x, \dot{x}) = n(x, \dot{x}) \, v(x, \dot{x}), \qquad (2.26)$$

which is essentially a product of two functions that could both serve as Lagrangians. The product rule for ordinary differentiation does not extend precisely to the variational derivative operator, but one does have:

$$\frac{\delta(nv)}{\delta x^i} = \frac{\delta n}{\delta x^i} v + n \frac{\delta v}{\delta x^i} - \frac{dn}{d\sigma} \frac{\partial v}{\partial \dot{x}^i} - \frac{\partial n}{\partial \dot{x}^i} \frac{dv}{d\sigma}. \qquad (2.27)$$

When one specializes the arbitrary parameter $\sigma$ back to arc-length $s$, the last term will drop out, and the Euler-Lagrange equations can be put into the form:

$$\frac{\delta n}{\delta x^i} = -n \frac{\delta v}{\delta x^i} + \frac{dn}{d\sigma} \frac{\partial v}{\partial \dot{x}^i}, \qquad (2.28)$$

or when written in terms of the partial derivatives instead of the variational ones:

$$\frac{\partial n}{\partial x^i} - \frac{d}{ds}\left(\frac{\partial n}{\partial \dot{x}^i}\right) = n g_{ij} \nabla_s \dot{x}^j + \frac{dn}{ds} \frac{\partial v}{\partial \dot{x}^i},$$

and when the derivatives of $v$ with respect to $\dot{x}^i$ from the previous calculations are substituted, one will have:

$$\frac{\partial n}{\partial x^i} - \frac{d}{ds}\left(\frac{\partial n}{\partial \dot{x}^i}\right) = g_{ij}(n\nabla_s \dot{x}^j + \frac{dn}{ds}\dot{x}^j),$$

which can be further simplified when one notes that the action of $\nabla_s$ on a scalar function is simply $d/ds$, and that the product rule applies to covariant differentiation, as well:

$$\frac{\partial n}{\partial x^i} - \frac{d}{ds}\left(\frac{\partial n}{\partial \dot{x}^i}\right) = g_{ij}\nabla_s(n\dot{x}^i). \tag{2.29}$$

In the case of an isotropic medium, for which $n$ will not depend upon $\dot{x}^i$, this can be expressed in the form:

$$\nabla_s(n\mathbf{t}) = \nabla n, \tag{2.30}$$

in which the symbol $\nabla$ on the right-hand side refers to the conventional gradient operator, this time. When the ambient space is Euclidian, one can use $d/ds$ for $\nabla_s$ and then expand the left-hand side, which will allow one to write this equation in the form:

$$\nabla n = \frac{dn}{ds}\mathbf{t} + (\kappa n)\mathbf{n} = (\mathbf{t}\, n)\,\mathbf{t} + (\mathbf{n}\, n)\,\mathbf{n}, \tag{2.31}$$

in which the notations $\mathbf{t}\, n$ and $\mathbf{n}\, n$ refer to the directional derivatives of $n$ in those directions. The final expression would be an identity, except that it is missing the contribution from the directional derivative of $n$ in the third spatial dimension. That suggests that the gradient of the index will always lie in the plane that is spanned by $\mathbf{t}$ and $\mathbf{n}$, namely, the osculating plane of the curve (i.e., the light ray). That is then an important property of the curves themselves.

An interesting aspect of equation (2.31) is that when one considers the normal component, one will get the equation:

$$\mathbf{n}\, n = \kappa n, \tag{2.32}$$

which takes the form of an eigenvalue equation for the index of refraction, with the curvature of the curve as the eigenvalue of the directional derivative operator in the normal direction. In any event, one can see that the curvature of the light ray that results in the medium will be driven primarily by the gradient of the index of refraction.

Typically, gradient-index optics [1] assumes that the ambient space is Euclidian, so the covariant derivative with respect to $s$ will reduce to the ordinary derivative with respect to $s$. One can then specialize the problem by imposing symmetries upon the medium and using the relevant systems of curvilinear coordinates. For instance, *axial* symmetry amounts to assuming Cartesian coordinates with an index that varies with only one dimension, such as altitude. In the example of the Earth's atmosphere, one sees that this can be regarded as an approximation to *spherical* symmetry when the radius is large. In the latter case, one employs spherical coordinates ($r$, $\theta$, $\psi$) and assume that the index varies with only $r$. A popular "toy model" for a gradient-index medium with spherical symmetry is the *Maxwell fish-eye* [**16-18**], for which the index of refraction varies with $r$ like:

$$n(r) = n_0 \left(1 + \frac{r^2}{a^2}\right)^{-1}. \tag{2.33}$$

in which $a$ is an empirical constant and $n_0$ is the index of refraction at the origin. As we shall see later, the Maxwell fish-eye relates to a number of static space-times that are considered in general relativity. A gradient-index medium can also have *radial* symmetry, for which one employs cylindrical coordinates ($r$, $\theta$, $z$) and assumes that the index varies with $r$.

**3. Geodesics of conformally-related metrics.** – A metric $\bar{g}$ is *conformally related* to another metric $g$ iff there is some strictly-positive spatial function $\Omega^2$ that makes:

$$\bar{g} = \Omega^2 g. \tag{3.1}$$

This has the effect of saying that when two tangent vectors **u** and **v** at any point in space are orthogonal under $g$ they will also be orthogonal under $\bar{g}$. More generally, a conformal change of metric will preserve all angles between lines through the origin. That is easy enough to verify when one uses the definition of the cosine between **u** and **v** :

$$\cos \angle (\mathbf{u}, \mathbf{v}) = \frac{g(\mathbf{u}, \mathbf{v})}{\|\mathbf{u}\| \|\mathbf{v}\|}, \qquad \|\mathbf{u}\|^2 = g(\mathbf{u}, \mathbf{u}), \qquad \|\mathbf{v}\|^2 = g(\mathbf{v}, \mathbf{v}). \tag{3.2}$$

Note that this definition can break down in spaces with metrics that are not definite, such as Minkowski space, in which non-zero vectors with zero norms can exist; indeed, that characterizes the tangent vectors to any light ray in space-time.

One can also think of this change of metric as being defined by a rescaling of the lengths of all tangent vectors by the function $\Omega$ :

$$\bar{g}(\mathbf{u}, \mathbf{v}) = g(\Omega \mathbf{u}, \Omega \mathbf{v}). \tag{3.3}$$

Such a transformation of each tangent space is referred to as a *homothety* (with its center at the origin).

In the case of isotropic optical media, one can use the index of refraction $n(x)$ as a scaling factor and define a conformal change of metric by:

$$\bar{g} = n^2 g. \tag{3.4}$$

Hence, one must generally consider two different metrics in optical problems: the Euclidian metric on the ambient space and the optical metric in the medium.

When one looks at the speed of a curve parameterization, one then sees that:

$$\bar{v}^2 = \bar{g}(\mathbf{v},\mathbf{v}) = n^2\, g(\mathbf{v},\mathbf{v}) = (n\,v)^2,$$

or

$$\bar{v} = n\,v. \tag{3.5}$$

This has the effect of also rescaling the arc-length of a curve to a new curve parameter $\bar{s} = \bar{s}(s)$ that relates to the old one by way of:

$$\frac{d\bar{s}}{ds} = \Omega, \qquad \text{i.e.,} \qquad d\bar{s} = \Omega\, ds. \tag{3.6}$$

Equation (3.5) has exactly the form that the Lagrangian for Fermat's principle took in (2.26), while (3.6) defines the optical path length that was defined in (2.22), at least for an isotropic optical medium. Hence, one can think of the metric of an isotropic optical medium as being conformally related to the ambient metric of the space that the medium is embedded in by way of using the index of refraction $n$ as the conformal scaling factor $\Omega$.

In order to see how a conformal change of metric will affect the geodesics of the original metric, one starts by recalculating the Christoffel symbols for $\bar{g}$:

$$\left\{\begin{matrix} i \\ jk \end{matrix}\right\}_\Omega = \tfrac{1}{2}\bar{g}^{il}(\partial_j \bar{g}_{lk} + \partial_k \bar{g}_{lj} - \partial_l \bar{g}_{jk}). \tag{3.7}$$

Since:

$$\bar{g}^{il} = \frac{1}{\Omega^2} g^{il}, \qquad \partial_i \bar{g}_{jk} = \partial_i \Omega^2 g_{jk} + \Omega^2 \partial_i g_{jk} = 2\Omega \partial_i \Omega\, g_{jk} + \Omega^2 \partial_i g_{jk},$$

one will get:

$$\left\{\begin{matrix} i \\ jk \end{matrix}\right\}_\Omega = \left\{\begin{matrix} i \\ jk \end{matrix}\right\} + \tau^i_{jk}, \tag{3.8}$$

in which we have introduced the *difference 1-form (with values in the Lie algebra of infinitesimal linear transformations):*

$$\tau^i_{jk} = \frac{1}{\Omega}(\partial_j \Omega\, \delta^i_k + \partial_k \Omega\, \delta^i_j - \partial^i \Omega\, g_{jk}), \tag{3.9}$$

which expresses the change in the connection 1-form under the conformal change of metric.

In order to relate the geodesics of the metric $\bar{g}$ to those of $g$, it is important to catch the subtlety that in order for the new equations to have the same form as the old ones, but with new variables, one must also change from the old arc-length $s$ to the new one $\bar{s}$ by way of $d\bar{s} = \Omega\, ds$. That is, the new form of the geodesic equations that one gets from the variational process will be:

$$0 = \bar{\nabla}_{\bar{s}}\, \dot{\bar{x}}^i = \frac{d^2 x^i}{d\bar{s}^2} + \left\{ \begin{matrix} i \\ j\, k \end{matrix} \right\}_\Omega \frac{dx^j}{d\bar{s}} \frac{dx^k}{d\bar{s}} = \frac{d^2 x^i}{d\bar{s}^2} + \left\{ \begin{matrix} i \\ j\, k \end{matrix} \right\} \frac{dx^j}{d\bar{s}} \frac{dx^k}{d\bar{s}} + \tau^i_{jk} \frac{dx^j}{d\bar{s}} \frac{dx^k}{d\bar{s}}.$$

In order to be able to compare this to the previous form, we must convert back to the previous arc-length $s$. The first term in the far right-hand equation will become:

$$\frac{d^2 x^i}{d\bar{s}^2} = \frac{ds}{d\bar{s}} \frac{d}{ds}\left(\frac{ds}{d\bar{s}} \frac{dx^i}{ds}\right) = \frac{1}{\Omega} \frac{d}{ds}\left(\frac{1}{\Omega} \frac{dx^i}{ds}\right) = -\frac{1}{\Omega^3} \frac{d\Omega}{ds} \frac{dx^i}{ds} + \frac{1}{\Omega^2} \frac{d^2 x^i}{ds^2},$$

while the second two get multiplied by $1/\Omega^2$ and the $\bar{s}$ in the derivatives gets replaced with $s$. After multiplying both sides by $\Omega^2$, that will make the new geodesic equation take the form:

$$0 = \nabla_s \frac{dx^i}{ds} - \frac{1}{\Omega} \frac{d\Omega}{ds} \frac{dx^i}{ds} + \tau^i_{jk} \frac{dx^j}{ds} \frac{dx^k}{ds}. \tag{3.10}$$

When one evaluates the last term on the right-hand side, while recalling that speed of the parameterization by $s$ was unity, one will get:

$$\tau^i_{jk} t^j t^k = \frac{1}{\Omega}(t^j \partial_j \Omega\, t^i + \frac{d\Omega}{ds} t^i - \partial^i \Omega) = \frac{1}{\Omega}(2\frac{d\Omega}{ds} t^i - \partial^i \Omega).$$

That will make the transformed geodesic equation (3.10) now take the form:

$$0 = \nabla_s t^i + \frac{1}{\Omega} \frac{d\Omega}{ds} t^i - \frac{1}{\Omega} \partial^i \Omega, \tag{3.11}$$

and after some rearranging:

$$\nabla_s (\Omega t^i) = \partial^i \Omega. \tag{3.12}$$

When one compares this to (2.30), one will see that when the conformal factor $\Omega$ is the index of refraction $n$, the optical metric that makes light rays into geodesics will be conformally related to the metric on the ambient space by way of $n^2$, although one must be careful to change the arc-length parameter accordingly. Therefore, in a sense, the difference between the length of a spatial curve and the time that it takes for a photon to traverse it is somewhat geometrically irrelevant as long as one makes a conformal change of metric accordingly.

One can recover a previous result, namely, (2.31), in a more geometrically-general form by recalling the Frenet-Serret equation $\nabla_s t^i = \kappa n^i$:

$$\nabla \Omega = (\mathbf{t}\, \Omega)\, \mathbf{t} + \kappa\, \Omega\, \mathbf{n}\,. \qquad (3.13)$$

We then have the more general:

**Theorem:**

1. *The osculating plane of the geodesic of the conformally-related metric $\bar{g}$ always contains the gradient of the conformal scale factor.*

2. *The curvature of that geodesic is an eigenvalue of the directional derivative in the normal direction with the conformal scale factor as its eigenfunction:*

$$\mathbf{n}\, \Omega = g\, (\nabla \Omega, \mathbf{n}) = \kappa\, \Omega\,. \qquad (3.14)$$

**4. Optical-mechanical analogies.** – Since the paths of light rays in an isotropic optical medium are conformally related to the metric on the ambient space, the question arises of whether one can sometimes represent the trajectories of point masses that are acted upon by external force fields as geodesics of metrics that are conformally related to the background metric, as well. As it turns out, the mechanical phenomena that can be modelled in that way are actually quite broad in scope.

*a. Jacobi's prescription.* – When a point mass $m$ moves in space under the influence of a time-invariant, conservative, external, force field $F\,(x) = F_i\,(x)\,dx^i$ that is derived from a potential function $U\,(x)$:

$$F = -\, dU\,, \qquad \text{i.e.,} \qquad F_i = \frac{\partial F}{\partial x^i}\,, \qquad (4.1)$$

its total energy:

$$E = T + U\,, \qquad T(x, \dot{x}) = \tfrac{1}{2} m v^2\,, \qquad v^2 = g_{ij}\, \dot{x}^i\, \dot{x}^j \qquad (4.2)$$

will then be constant.

Now, consider de Maupertuis's form of the least-action principle, which defines the action functional on a spatial path $x\,(t)$ by:

$$S\,[x\,(t)] = \int_{x(t)} p = \int_{x(t)} p_i\, dx^i = \int_{t_0}^{t_1} p_i(t)\, \dot{x}^i(t)\, dt = \int_{t_0}^{t_1} 2T(x(t), \dot{x}(t))\, dt\,, \qquad (4.3)$$

in which:

$$p = p_i\,(t)\, dx^i, \qquad p_i = m\, g_{ij}\, \dot{x}^j \qquad (4.4)$$

is the linear momentum of the moving mass.

Since:

$$\int_{t_0}^{t_1} 2T\, dt = m \int_{t_0}^{t_1} v^2(x(t), \dot{x}(t))\, dt\,, \qquad (4.5)$$

and $m$ is assumed to be constant, the extremal curves of the de Maupertuis action prove to the purely kinematical geodesics that are defined by the ambient metric $g$ and its Levi-Civita connection.

What Jacobi suggested in his lectures on mechanics [**19**] is that if one assumes conservation of energy before forming the action functional, rather than deriving it as a consequence, then one can *replace* the expression $2T$ with $2(E - U)$ and get an action functional:

$$S[x(t)] = \int_{t_0}^{t_1} 2(E-U)\, dt \tag{4.6}$$

that gives the equations of motion of the mass moving under the influence of a time-invariant, conservative force field. The reason that we emphasize the verb "replace" is that $T$ is a function of both position and velocity, while $E - U$ is a function of only position, so one cannot strictly equate one to the other.

When one solves the second of equations (4.2) for $v$, while using the latter replacement:

$$v = \sqrt{\frac{2(E-U)}{m}} = \frac{ds}{dt}, \tag{4.7}$$

one will see that:

$$dt = \sqrt{\frac{m}{2(E-U)}}\, ds, \qquad \text{so} \qquad 2(E-U)\, dt = \sqrt{2m(E-U)}\, ds. \tag{4.8}$$

Hence, we can define a conformal transformation of the ambient metric $g$:

$$\bar{g} = \Omega^2 g, \qquad \Omega^2 = 2m(E-U) \tag{4.9}$$

that makes (4.6) take the form of an arc-length functional on the curve when the arc-length is rescaled to:

$$d\bar{s} = \Omega\, ds. \tag{4.10}$$

From what we saw above in (3.11), the extremals of the new arc-length functional will take the form of geodesics of the metric $\bar{g}$, so one will have:

$$0 = \bar{\nabla}_s t^i = \nabla_s t^i + \frac{1}{2\Omega^2}\left(\frac{d\Omega^2}{ds} t^i - \partial^i \Omega^2\right), \tag{4.11}$$

or

$$\nabla_s t^i = \frac{1}{2\Omega^2}\left(\partial^i \Omega^2 - \frac{d\Omega^2}{ds} t^i\right). \tag{4.12}$$

With the substitution for $\Omega^2$ that is given in the second of equations (4.9), one will get:

$$\frac{d\Omega^2}{ds} = -2m\frac{dU}{ds}, \qquad \partial^i \Omega^2 = -2m\,\partial^i U, \tag{4.13}$$

so the geodesic equations will take the form:

$$\nabla_s t^i = \frac{1}{2(E-U)}(\frac{dU}{ds}t^i - \partial^i U), \tag{4.14}$$

but since $2(E - U) = m v^2$, one can put this into the form:

$$mv^2 \nabla_s t^i = \frac{dU}{ds}t^i - \partial^i U. \tag{4.15}$$

From this, the form that the theorem at the end of the last section takes is now:

$$F = -\nabla U = -(\mathbf{t}U)\mathbf{t} + (\kappa mv^2)\mathbf{n}, \tag{4.16}$$

namely, *the osculating plane of the geodesic always contains the gradient of the potential function – i.e., the force vector – and the curvature of that geodesic is an eigenvalue of the normal derivative operative with the potential function as its eigenfunction.*

If $F^n$ is the normal component of $-\nabla U$ then:

$$F^n = \kappa mv^2 = \frac{mv^2}{r}, \tag{4.17}$$

in which $r$ is the radius of curvature of the geodesic at each point. Hence, the curvature is basically the centrifugal force of motion.

We now change the parametrization of the curve from arc-length to time $t$. With the substitutions:

$$\frac{dx^i}{ds} = \frac{1}{v}\frac{dx^i}{dt}, \qquad \frac{d^2 x^i}{ds^2} = \frac{1}{v}\frac{d}{dt}\left(\frac{1}{v}\frac{dx^i}{dt}\right) = -\frac{1}{v^3}\frac{dv}{dt}\dot{x}^i + \frac{1}{v^2}\ddot{x}^i, \tag{4.18}$$

the geodesic equation (4.15) will initially take the form:

$$m\nabla_t \dot{x}^i - \frac{m}{v}\frac{dv}{dt}\dot{x}^i = -\partial^i U + \frac{1}{v^2}\frac{dU}{dt}\dot{x}^i,$$

but since $U = E - \frac{1}{2} m v^2$, one will have:

$$\frac{1}{v^2}\frac{dU}{dt} = -\frac{m}{2v^2}\frac{dv^2}{dt} = -\frac{m}{v}\frac{dv}{dt},$$

and all that will remain is:

$$m \nabla_t \dot{x}^i = -\partial^i U = F^i, \qquad (4.19)$$

which has the familiar "$F = ma$" form from Newtonian mechanics, with a more general definition of the acceleration.

Interestingly, one can start with Fermat's principle – i.e., the elapsed-time functional – instead of de Maupertuis and still get the same geodesics. Recall the relationship in (4.8), and express $dt$ in "optical" form:

$$dt = n(x)\,ds, \qquad (4.20)$$

in which we have introduced an effective index of refraction for the ambient space that is due to the potential energy function:

$$n(x) = \sqrt{\frac{m}{2(E - U(x))}}. \qquad (4.21)$$

If one follows through with the calculations that give the geodesic equations for the elapsed-time functional:

$$T[x(s)] = \int_{x(s)} dt = \int_{x(s)} n(x(s))\,ds = \int_{x(s)} \sqrt{\frac{m}{2[E - U(x(s))]}}\,ds \qquad (4.22)$$

then one will find that one gets the same equations as before, namely, (4.15) for the geodesics when they are parameterized by arc-length. Hence, one will also get the same equations relative to the time parameter, as well.

That tends to suggest that there is a close analogy between the presence of a continuously-varying index of refraction in an optical medium and a force potential function in space.

*b. Levi-Civita's conversion.* – In a paper [20] that Tulio Levi-Civita published in 1917 on the subject of Einsteinian statics, he addressed the way that one might obtain the geodesics of point masses in three-dimensional space from the geodesics in four-dimensional space-time. In particular, he addressed the form that the spatial arc-length element might take.

In a static space-time, not only is the Lorentzian metric assumed to be time-invariant (most simply, that means that its components do not depend upon time in the chosen Lorentzian frame), but one also has a time+space splitting of the tangent spaces to space-time that allows one to express the metric in the form:

$$d\bar{s}^2 = c_0^2\,d\tau^2 = (dx^0)^2 - g_{ij}\,dx^i\,dx^j, \qquad (4.23)$$

in which the four-dimensional arc-length parameter $\bar{s}$ becomes proportional to proper-time $\tau$. Thus, for a four-velocity vector field $\mathbf{u}(\tau) = dx/d\tau$ along a world-line $x(\tau)$ that is parameterized by proper-time, whose temporal component is $u^0 = c_0\,\gamma$ and whose spatial components are $\gamma\,v^i$, with:

$$\gamma = \left(1 - \frac{v^2}{c_o}\right)^{-1/2}, \tag{4.24}$$

one will find that the speed of the parameterization $\bar{s}$ with respect to $\tau$ takes the form:

$$\frac{d\bar{s}}{d\tau} = \|\mathbf{u}\| = \gamma(c_0^2 - v^2)^{1/2}, \quad \text{i.e.,} \quad d\bar{s} = \gamma(c_0^2 - v^2)^{1/2}\,d\tau. \tag{4.25}$$

Since one also has that:

$$\gamma = \frac{dt}{d\tau}, \tag{4.26}$$

one can say that:

$$d\bar{s} = (c_0^2 - v^2)^{1/2}\,dt. \tag{4.27}$$

Since one also has that:

$$v = \frac{ds}{dt}, \tag{4.28}$$

one can conclude that:

$$d\bar{s} = \frac{1}{v}(c_0^2 - v^2)^{1/2}\,ds = \left(\frac{c_0^2}{v^2} - 1\right)^{1/2}\,ds. \tag{4.29}$$

Hence, one has made a conformal change of the spatial metric by the scale factor:

$$\Omega = \left(\frac{c_0^2}{v^2} - 1\right)^{1/2}. \tag{4.30}$$

What Levi-Civita proceeded to do was to allow the speed of light to vary with position, and thus replaced the constant $c_0$ with the spatial function $c(x)$. In the space-time arc-length functional:

$$\bar{s}[x(\tau)] = \int_{x(\tau)} d\bar{s} = \int_{x(t)} (c^2 - v^2)^{1/2}\,dt, \tag{4.31}$$

one could then define the Lagrangian function by:

$$\mathcal{L}(x,\dot{x}) = \begin{cases} (c^2 - v^2)^{1/2} & c \geq v, \\ (v^2 - c^2)^{1/2} & c < v. \end{cases} \tag{4.32}$$

Rather than specify the corresponding geodesic equations, he then proceeded to take advantage of the fact that the time-invariance of $\mathcal{L}$ allows one to define a constant of the motion. In order to find it, one first calculates the total time derivative of $\mathcal{L}$ with respect to $t$ and applies the Euler-Lagrange equations to the particular derivatives with respect to $x^i$:

$$\frac{d\mathcal{L}}{dt} = \frac{\partial \mathcal{L}}{\partial t} + \frac{\partial \mathcal{L}}{\partial x^i}\dot{x}^i + \frac{\partial \mathcal{L}}{\partial \dot{x}^i}\ddot{x}^i = \frac{\partial \mathcal{L}}{\partial t} + \frac{d}{dt}\left(\frac{\partial \mathcal{L}}{\partial \dot{x}^i}\right)\dot{x}^i + \frac{\partial \mathcal{L}}{\partial \dot{x}^i}\ddot{x}^i = \frac{\partial \mathcal{L}}{\partial t} + \frac{d}{dt}\left(\frac{\partial \mathcal{L}}{\partial \dot{x}^i}\dot{x}^i\right) = \frac{d}{dt}\left(\frac{\partial \mathcal{L}}{\partial \dot{x}^i}\dot{x}^i\right),$$

in which we have employed the time-invariance of $\mathcal{L}$ in the last step. That will make:

$$0 = \frac{d}{dt}\left(\frac{\partial \mathcal{L}}{\partial \dot{x}^i}\dot{x}^i - \mathcal{L}\right), \tag{4.33}$$

so the expression in parentheses will be constant in time, and we call that constant $C_0$:

$$C_0 = \frac{\partial \mathcal{L}}{\partial \dot{x}^i}\dot{x}^i - \mathcal{L}. \tag{4.34}$$

Explicit calculation using the expression for $\mathcal{L}$ above gives:

$$C_0 = -\frac{c^2}{\sqrt{\pm(c^2 - v^2)}}, \tag{4.35}$$

and solving that for $v$ gives:

$$v = c\left(1 \mp \frac{c^2}{C_0^2}\right)^{1/2}. \tag{4.36}$$

Substituting this in the Lagrangian $\mathcal{L}$ makes it take the form:

$$\mathcal{L} = \mp \frac{c^2}{C_0}. \tag{4.37}$$

In order to derive Levi-Civita's result, one must replace this with:

$$\mathcal{L} + C_0 = C_0\left(1 \mp \frac{c^2}{C_0^2}\right), \tag{4.38}$$

which will not change the equations of the geodesics.

The relationship between the space-time arc-length $d\bar{s}$ and the spatial arc-length $ds$ becomes:

On the geodesics of gradient-index optical metrics and the optical-mechanical analogy.                    20$$d\bar{s} = \frac{\mathcal{L}+C_0}{v} ds = \frac{1}{C_0}\left(\frac{1}{c^2} \mp \frac{1}{C_0^2}\right)^{1/2} ds, \tag{4.39}$$

which is the form that Levi-Civita obtained for this, except that he suppressed the constant multiple. Indeed, he defined the potential function $U$ by:

$$2U = \frac{1}{c^2} \mp \frac{1}{C_0^2}. \tag{4.40}$$

Thus, one comes back to the Jacobi prescription by treating the conformal scale factor in front of $ds$ as if it were related to the square root of a potential function $U$. In fact, in the next section, Levi-Civita showed that with:

$$c(x) = c_0[1 + \eta(x)], \tag{4.41}$$

if one assumes that $\eta$ is sufficiently small then one could show that the function $\eta(x)$ satisfied an equation of Poisson type that justified treating it (or rather $c$) as a gravitational potential. This concept had been explored previously by Max Abraham [21] and B. Caldonazzo [22]. According to Levi-Civita, it is also how Einstein settled upon the expression that he used for the proportionality constant in his field equations for gravitation.

Before Erwin Schrödinger arrived at his approach to wave mechanics in terms of the eigenvalues of Hermitian operators, he was pursuing the analogy between the relationship between wave optics and geometrical optics and the relationship between wave mechanics and point mechanics. In some unpublished notes from around 1918 [23], he took the Jacobi prescription as his starting point and then went on to examine the Hamilton-Jacobi picture that it led to.

  *c. "F = ma" optics.* – When one reconsiders the basic equation (2.30) for the geodesics of an inhomogeneous, isotropic optical medium, one sees that there is a partial analogy with the equations of motion in Newtonian mechanics for a point mass that moves in a conservative external force field. In order to make the "$F = ma$" analogy closer, Evans and Rosenquist [24] chose to first set $m = 1$ and then introduced another curve parameter $\sigma$ that they called the "stepping parameter," which related to the arc-length $s$ by way of:

$$ds = n\, d\sigma, \quad \text{i.e.,} \quad \frac{ds}{d\sigma} = n. \tag{4.42}$$

When one multiplies both sides of (2.30) by $n$, one will get:

$$n\frac{d}{ds}\left(n\frac{dx}{ds}\right) = \frac{d^2x}{d\sigma^2}, \qquad n\nabla n = \tfrac{1}{2}\nabla n^2,$$

and therefore:

$$\frac{d^2x}{d\sigma^2} = \nabla\left(\tfrac{1}{2}n^2\right). \tag{4.43}$$

Hence, if one regards the expression on the left-hand side as a type of acceleration then one can regard $-\tfrac{1}{2}n^2$ as a potential function for a conservative external force. One can then summarize the resulting analogy in the following table:

Table 1. – The optical-mechanical analogy.

| Quantity | Mechanics | Optics |
| --- | --- | --- |
| Position | $x(t)$ | $x(\sigma)$ |
| Curve parameter | time $t$ | stepping parameter $\sigma$ |
| Speed of parameterization | $v = ds/dt$ | $n = ds/d\sigma$ |
| Potential function | $U(x)$ | $-\tfrac{1}{2}n^2$ |
| Mass | $m$ | 1 |
| Kinetic energy | $\tfrac{1}{2}mv^2$ | $\tfrac{1}{2}n^2$ |
| Total energy | $\tfrac{1}{2}mv^2 + U$ | 0 |
| Action functional | $\int (\tfrac{1}{2}mv^2 - U)\,dt$ | $\int n^2\,d\sigma$ |

In particular, one sees that the total energy of motion with this parameterization of the light ray is zero, because the kinetic energy proves to be minus the potential energy. That basically comes down to the fact that the curve parameter $\sigma$ gets rescaled by an external function (namely, $n$) in order to get the arc-length.

*d. Hydrodynamical models.* – Lichnerowicz [**25**] defined a fluid medium to be *holonomic* when its streamlines could be obtained from the geodesics of a metric that is conformally related to the ambient space metric. He called the conformal scaling factor the "index" of the fluid.

One also finds that the definition of the *dynamic pressure* for a fluid of mass density $\rho(x)$ that moves with a flow velocity **v** takes the form:

$$p_d = \tfrac{1}{2}\rho v^2, \tag{4.44}$$

which can also be regarded as the kinetic energy density of the moving fluid. In this case, since it is the square of the speed that is involved, one must use $\Omega = \sqrt{\rho}$ as the conformal factor, rather than $\rho$ itself.

When one defines the conformal transformation $g = \rho\,\delta$ of the basic Euclidian metric on $\mathbb{R}^n$, the covariant derivative of the flow velocity vector field will now take the form:

$$\bar{\nabla}_t \dot{x}^i = \frac{d\dot{x}^i}{dt} + \tau^i_{jk} \dot{x}^j \dot{x}^k, \tag{4.45}$$

in which:

$$\tau^i_{jk} = \frac{1}{2\rho} (\partial_j \rho \, \delta^i_k + \partial_k \rho \, \delta^i_j - \partial_i \rho \, \delta_{jk}), \tag{4.46}$$

so

$$\tau^i_{jk} \dot{x}^j \dot{x}^k = \frac{1}{\rho} \left( \frac{d\rho}{dt} \dot{x}^i - \tfrac{1}{2} v^2 \partial^i \rho \right) = -\frac{v^2}{2\rho} \partial^i \rho, \tag{4.47}$$

in which we have use conservation of mass (i.e., the continuity equation) to eliminate $d\rho / dt$.

Hence, the geodesics of the deformed matrix will obey the equation:

$$\rho \frac{d\dot{x}^i}{dt} = \tfrac{1}{2} v^2 \partial^i \rho. \tag{4.48}$$

which has the "$F = ma$" form if one regards the force density that acts upon the fluid as something that is proportional to the gradient of the mass density. Hence, the geodesics will be straight lines iff the mass density is uniform. Of course, the equations of motion of the fluid (e.g., the Euler equations) couple the left-hand side of (4.48) to the other force densities than the one on the right-hand side (e.g., pressure gradient, gravity), so the actual path-lines or streamlines will not generally be the same as the solutions to the last equation.

**5. Gradient-index optics in general relativity.** – The applications of gradient-index optics to general relativity that we shall discuss will be confined to the way that the aforementioned gravitational lensing and Gordon's optical metric define effective indices of refraction in strong gravitation fields and optical media in a state of the relative motion to an observer, respectively.

*a. Gravitational lensing.* – Any metric that is expressed in general coordinates in the form:

$$g = g_{00} c_0^2 dt^2 + 2 g_{0i} c_0 \, dt \, dx^i + g_{ij} \, dx^i \, dx^j \tag{5.1}$$

can be converted into the form:

$$g = c_0^2 dt^2 - n^2 \gamma_{ij} \, dx^i \, dx^j \tag{5.2}$$

by changing to another time-like observer whose time-line has the covelocity covector field:

$$u = c_0 \, dt + v, \quad \text{so} \quad c_0 \, dt = u - v, \tag{5.3}$$

in which $v = v_i \, dx^i$ is a suitable spatial covelocity.

Substitution of the latter expression in (5.1) gives:

$$g_{00}(u-v)^2 + 2g_{0i}(u-v)dx^i + g_{ij}dx^i dx^j$$
$$= g_{00}u^2 - 2(g_{00}v_i + g_{0i})u\, dx^i - (2g_{0i}v_j + g_{00}v_i v_j + g_{ij})dx^i dx^j.$$

One can make the time-space component of $g$ vanish by setting:

$$v_i = -\frac{g_{0i}}{g_{00}}, \tag{5.4}$$

which will make:

$$2g_{0i}v_j + g_{00}v_i v_j = -\frac{g_{0i}g_{0j}}{g_{00}}.$$

Thus, when one defines the spatial metric by:

$$ds^2 = \gamma = \gamma_{ij}dx^i dx^j, \qquad \gamma_{ij} = g_{ij} - \frac{g_{0i}g_{0j}}{g_{00}}, \tag{5.5}$$

one will have $g$ in the form:

$$g = g_{00}u^2 - \gamma_{ij}dx^i dx^j. \tag{5.6}$$

Since the associations (5.3), (5.4) and (5.5) are invertible, one can always find an observer $u$ that puts $g$ into the form (5.6).

If $u$ takes the form $c_0\, d\bar{t}$ for some other time coordinate $\bar{t}$ then one can put $g$ into the form:

$$g = g_{00}(c_0^2\, d\bar{t}^2 - n^2\gamma), \qquad n \equiv (c_0\, g_{00})^{-1}, \tag{5.7}$$

which is conformal to (5.2), so the two metrics will define the same light cones.

Many of the Lorentzian metrics that are used by general relativity have the form (5.6) with $\gamma = \Omega^2\delta$, which is typical of *static*, *spatially isotropic* space-times when the components of the metric are also independent of time, although generally the Euclidian metric $\delta = \delta_{ij}dx^i dx^j$ gets expressed in a coordinate system that is adapted to the nature of the gravitating body, such as spherical coordinates for most celestial bodies. For instance, the general form for a metric that is produced by a static, spherically-symmetric gravitating body will be the Schwarzschild metric [6]:

$$g = \left(1-\frac{R_s}{r}\right)^2\left(1+\frac{R_s}{r}\right)^{-2}c_0^2\, dt^2 - \left(1+\frac{R_s}{r}\right)^4\delta_{ij}dx^i dx^j. \tag{5.8}$$

in which the constant $R_s$, which is called the *Schwarzschild radius* of the massive body whose mass is $M$, is defined by:

$$R_s = \frac{2GM}{c_0^2}, \tag{5.9}$$

and $G$ is Newton's gravitational constant. This solution to the Einstein field equations for gravitation is quite useful in describing celestial bodies, such as moons, planets, and stars that are not too dense.

As long as one is only dealing with light cones, one can define a conformally-related Lorentzian metric by factoring out the $g_{00}$ component in (5.8):

$$\bar{g} = c_0^2\, dt^2 - n^2 ds^2 \tag{5.10}$$

that makes:

$$g = \left(\frac{1 - R_s/r}{1 + R_s/r}\right)^2 \bar{g}, \tag{5.11}$$

and still be dealing with the same light cones.

Meanwhile, we are defining the equivalent index of refraction to be:

$$n(r) = \frac{\left(1 + \dfrac{R_s}{r}\right)^3}{1 - \dfrac{R_s}{r}}. \tag{5.12}$$

For weak gravitational fields $r \gg R_s$, and the exterior refractive index can be approximated by [**9**]:

$$n(r) \approx e^{R_s/r} = \exp\left(-\frac{2}{c^2} U(r)\right), \tag{5.13}$$

in which:

$$U(r) = -\frac{GM}{r} \tag{5.14}$$

is the conventional Newtonian potential.

Some other examples of equivalent indices of refraction are worked out in [**26**]. For instance, in isotropic coordinates, the *de Sitter space-time* is characterized by the metric:

$$g = \left(\frac{1 - \tfrac{1}{12}\Lambda r^2}{1 + \tfrac{1}{12}\Lambda r^2}\right)^2 c_0^2\, dt^2 - (1 + \tfrac{1}{12}\Lambda r^2)^{-2} ds^2, \tag{5.15}$$

in which $\Lambda$ is the cosmological constant.

That spacetime then has an equivalent index of refraction that is defined by:

$$n(r) = (1 - \tfrac{1}{12}\Lambda r^2)^{-1}, \tag{5.16}$$

and when $\Lambda = -K < 0$ (i.e., an open spacetime), that will take the form of the Maxwell fish-eye when one sets:

$$n_0 = 1, \qquad a^2 = \frac{12}{K}. \tag{5.17}$$

The *Robertson-Walker metric*, which characterizes an expanding universe, has a metric that is expressed in isotropic coordinates in the form:

$$g = c_0^2 \, dt^2 - \left(\frac{R(t)}{1 + \tfrac{1}{4} k r^2}\right)^2 ds^2, \tag{5.18}$$

which then admits the equivalent index of refraction:

$$n(r) = \frac{R(t)}{1 + \tfrac{1}{4} k r^2}. \tag{5.19}$$

This also takes the form of the Maxwell fish-eye, but with a time-varying value for the index of refraction at the origin:

$$n_0(t) = R(t), \qquad a^2 = \frac{4}{k}. \tag{5.20}$$

*b. Gordon's optical metric* [**10**]. – What Gordon was looking into was the question of how the relative motion of a polarizable electromagnetic medium would affect propagation of electromagnetic waves in the medium, and in particular, whether one could define an effective metric to account for the change.

The electromagnetic field strength 2-form $F$ and the electromagnetic excitation bivector field $\mathfrak{H}$, which can be expressed in the time+space form:

$$F = \frac{1}{c_0} u \wedge E + B, \qquad \mathfrak{H} = \frac{1}{c_0} \mathbf{u} \wedge \mathbf{D} + \mathbf{H}, \tag{5.21}$$

in which $u$ is the covelocity 1-form for a time-like observer, $\mathbf{u}$ is their four-velocity vector field, the electric field strength $E$ is a spatial covector field, the magnetic flux density $B$ is a spatial 2-form, the electric excitation (or displacement) $\mathbf{D}$ is a spatial vector field, and the magnetic field strength $\mathbf{H}$ is a spatial bivector field ([1]).

---

([1])  It is important to notice something about the definitions of $F$ and $\mathfrak{H}$ that is not obvious in the usual relativistic discussion of electromagnetism, which are confined to the classical vacuum: The way that the 2-form and bivector field are constructed mixes "kinematical" fields (i.e., field strengths) with "dynamical" ones (i.e., excitations).

When one evaluates the 2-form $F$ on the bivector field $\mathfrak{H}$, one will get the scalar field:

$$F(\mathbf{H}) = \frac{1}{c_0^2} E(\mathbf{D}) + B(\mathbf{H}), \qquad (5.22)$$

which takes the form of the Lagrangian density for the field in that medium. Although the second term appears to differ in sign from the conventional expression, that is only because one is contracting a spatial 2-form on a spatial bivector field, and when one associates them with a spatial vector field and covector field, respectively, by way of the spatial volume element (which will be discussed below), the sign will change.

In a coordinate system that is adapted to the splitting of the tangent bundle $T(M) = [\mathbf{u}] \oplus \Sigma(M)$ that $\mathbf{u}$ defines (so $\mathbf{u} = c_0\, \partial_0$, $u = c_0\, dx^0$, and the vector fields $\partial_i$, $i = 1, 2, 3$ span the "rest space" $\Sigma_x(M)$ at each point), the fields will take the form:

$$F = \tfrac{1}{2} F_{\mu\nu}\, dx^\mu \wedge dx^\nu = \frac{1}{c_0} E_i\, dx^0 \wedge dx^i + \tfrac{1}{2} B_{ij}\, dx^i \wedge dx^j, \qquad (5.23)$$

$$\mathfrak{H} = \tfrac{1}{2} \mathfrak{H}^{\mu\nu}\, \partial_\mu \wedge \partial_\nu = \frac{1}{c_0} D^i\, \partial_0 \wedge \partial_i + \tfrac{1}{2} H^{ij}\, \partial_i \wedge \partial_j. \qquad (5.24)$$

One can obtain the spatial fields by projection operators onto the temporal $[\mathbf{u}]_x$ and spatial subspaces $\Sigma_x$ of $\Lambda_{x,2} M$ and $\Lambda_x^2 M$ that take the forms of interior products:

$$E = \frac{1}{c_0} i_{\mathbf{u}} F, \quad \mathbf{H} = \frac{1}{c_0^2} i_u (\mathbf{u} \wedge \mathfrak{H}), \quad \mathbf{D} = \frac{1}{c_0} i_u \mathfrak{H}, \quad B = \frac{1}{c_0^2} i_u (u \wedge F). \qquad (5.25)$$

In an coordinate system that is adapted to the splitting, they take the local component forms:

$$E_i = F_{0i}, \qquad B_{ij} = F_{ij}, \qquad D^i = \mathfrak{H}^{0i}, \qquad H^{ij} = \mathfrak{H}^{ij}. \qquad (5.26)$$

Typically, the spatial 2-form $B$ and bivector field $\mathbf{H}$ are represented as spatial (axial) vectors, which is accomplished by using the spatial dual operator $\#_s$ that is defined by the spatial volume element:

$$V_s = dx^1 \wedge dx^2 \wedge dx^3 = \frac{1}{3!} \varepsilon_{ijk}\, dx^i \wedge dx^j \wedge dx^k. \qquad (5.27)$$

The invertible linear operator $\#_s : \Lambda_1 \Sigma_x (= \Sigma_x) \to \Lambda^2 \Sigma_x$ then takes a vector $\mathbf{v}$ to the 2-form:

$$\#_s \mathbf{v} = i_\mathbf{v} V_s = \tfrac{1}{2} (\varepsilon_{ijk} v^k)\, dx^i \wedge dx^j. \qquad (5.28)$$

This can be extended to an operator $\#_s : \Lambda_2 \Sigma_x \to \Lambda^1 \Sigma_x (= \Sigma_x^*)$ by extending the interior product operator to bivectors. The 1-form that is dual to a bivector field $\mathbf{A}$ will then be:

$$\#_s \mathbf{A} = i_\mathbf{A} V_s = \tfrac{1}{2} (\varepsilon_{ijk} A^{jk}) dx^i . \tag{5.29}$$

Hence, one can define a spatial vector field $\mathbf{B}$ and a spatial 2-form $H$ such that:

$$B = \#_s \mathbf{B} = \tfrac{1}{2}(\varepsilon_{ijk} B^k) dx^j \wedge dx^k , \qquad \mathbf{H} = \#_s^{-1} H = \tfrac{1}{2}(\varepsilon^{ijk} H_{jk}) \partial_i . \tag{5.30}$$

One can now introduce *electromagnetic constitutive laws*, which take the form of invertible functional relationships:

$$\mathbf{D} = \mathbf{D}\ (E, \mathbf{H}), \qquad B = B\ (E, \mathbf{H}) . \tag{5.31}$$

(At least, when the medium is "non-dispersive," which would make the operators in question include integral operators.)

When there is no cross-coupling of the fields (such as in the Faraday effect and optical activity), they reduce to the forms:

$$\mathbf{D} = \varepsilon\ (E), \qquad B = \mu\ (\mathbf{H}) , \tag{5.32}$$

in which the operators $\varepsilon$ and $\mu$ are referred to as the *dielectric strength* and *magnetic permittivity* of the medium, respectively. In optics, $\varepsilon$ can be a linear or nonlinear operator, but generally $\mu$ is a linear operator takes the component form $\mu_{ij} = \mu_0\ g_{ij}$, where $\mu_0$ is constant (optical media are not usually magnetizable), and $g_{ij}$ is the spatial metric, such as the Euclidian one, for which $g_{ij} = \delta_{ij}$. That would make the medium magnetically linear, homogeneous, and *isotropic*; typically, $\mu_0$ is set to 1. Similarly, it can be electrically linear and isotropic, and in that case $\varepsilon$ would take the component form $\varepsilon^{ij} = \varepsilon\ g^{ij}$; in the homogeneous case, $\varepsilon$ would be a constant.

One can define a spatial vector field $\mathbf{E}$ that is dual to the covector field $B$ by way of the spatial metric (so its components as $E^i = g^{ij} E_j$). For an electrically and magnetically-isotropic medium, that would make:

$$\mathbf{D} = c_0^2\ \varepsilon\ \mathbf{E}, \quad \mathbf{B} = \mu\ \mathbf{H} . \tag{5.33}$$

Hence, in such a case, the definition of $\mathfrak{H}$ can be given the form:

$$\mu\ \mathfrak{H} = c_0\ \varepsilon\mu\ \mathbf{u} \wedge \mathbf{E} + \mathbf{B} . \tag{5.34}$$

In the relativistic theory of electromagnetism, one generally deals with the "classical" vacuum, which is defined to be non-dispersive, linear, isotropic, and homogeneous, so its dielectric properties and magnetic permeability will be described by two constants $\varepsilon_0$ and $\mu_0$. Since constancy and isotropy already sound like relativistically-non-invariant concepts, one typically only encounters the two constants in the form of the speed of propagation of light *in vacuo*:

$$c_0 = \frac{1}{\sqrt{\varepsilon_0 \mu_0}}, \tag{5.35}$$

which is, by hypothesis, a Lorentz invariant.

The way that one associates $F$ with $\mathfrak{H}$ is by way of the linear isomorphism of $\Lambda^2 M$ with $\Lambda_2 M$ that is defined by the Lorentzian metric $g$ (also known as "raising both indices."). That is, one sets:

$$\mathfrak{H}^{\mu\nu} = \tfrac{1}{2}(g^{\mu\kappa} g^{\nu\lambda} - g^{\nu\kappa} g^{\mu\lambda}) F_{\mu\nu}. \tag{5.36}$$

The basic problem that Gordon was addressing in [**10**] was whether one could define a space-time metric $\gamma$ that would make the constitutive law for isotropic media take the form (5.36). One first solves (5.34) for **B** and substitutes:

$$\mathbf{E} = \frac{1}{c_0} i_u \mathbf{F}, \tag{5.37}$$

in which **F** is the metric dual to $F$ using $g$. That will make:

$$\mathbf{B} = \mu\,\mathfrak{H} - \varepsilon\mu\,\mathbf{u} \wedge i_u \mathbf{F}. \tag{5.38}$$

Solving the equation for $F$ for $B$, while substituting the definition of $E$ and raising both indices will give:

$$\mathbf{B} = \mathbf{F} - \frac{1}{c_0^2}\mathbf{u} \wedge i_u \mathbf{F} = \mathbf{F} - \varepsilon_0 \mu_0\,\mathbf{u} \wedge i_u \mathbf{F} \tag{5.39}$$

Equating the two expressions for **B** will allow one to express $\mu\,\mathfrak{H}$ in terms of **F**:

$$\mu\,\mathfrak{H} = \mathbf{F} + (\varepsilon\mu - \varepsilon_0 \mu_0)\,\mathbf{u} \wedge i_u \mathbf{F}, \tag{5.40}$$

which can be written in components:

$$\mu\,\mathfrak{H}^{\mu\nu} = [g^{\mu\kappa} g^{\nu\lambda} + (\varepsilon\mu - \varepsilon_0 \mu_0)\,u^\mu u^\kappa g^{\nu\lambda}] F_{\kappa\lambda} \qquad [\kappa\lambda], \tag{5.41}$$

where the notation $[\kappa\lambda]$ means that right-hand side must be anti-symmetrized with respect to those two indices.

Now, the expression in square brackets can be factored with respect to the antisymmetric product into the form:

$$[g^{\mu\kappa} + (\varepsilon\mu - \varepsilon_0 \mu_0)\,u^\mu u^\kappa]\,[g^{\nu\lambda} + (\varepsilon\mu - \varepsilon_0 \mu_0)\,u^\nu u^\lambda] \qquad [\kappa\lambda],$$

since the product $u^\mu u^\kappa u^\nu u^\lambda$ is completely symmetric in its indices and will thus vanish when one antisymmetrizes it with respect to $\kappa$ and $\lambda$.

Hence, one concludes with the Gordon definition of the "optical" metric:

$$\gamma^{\mu\nu} = g^{\mu\nu} + (\varepsilon\mu - \varepsilon_0\mu_0)\, u^\mu u^\nu. \tag{5.42}$$

In a frame that is adapted to **u**, $u^0 = c_0$, $u^i = 0$, the only component of $g$ that will be affected will be the time-time component:

$$\gamma^{00} = g^{00} + (\varepsilon\mu - \varepsilon_0\mu_0)\, c_0^2 = g^{00} + \left(\frac{c_0}{c}\right)^2 - 1, \tag{5.43}$$

and in Minkowski space, for which $g^{00} = 1$, that will make:

$$\gamma^{00} = \frac{\varepsilon\mu}{\varepsilon_0\mu_0} = \left(\frac{c_0}{c}\right)^2 = \frac{1}{n^2} \tag{5.44}$$

Thus, the Gordon metric effectively includes the index of refraction of the medium.

**6. Conclusions.** – In summary, we can say:

1. The spatial paths of light in an isotropic optical medium whose index of refraction has a non-vanishing gradient are the geodesics of a spatial metric $\bar{g}$ that is conformally related to the metric $g$ on the ambient space by way of the square of the index of refraction $n$:

$$\bar{g} = n^2\, g. \tag{6.1}$$

3. The geodesics of conformally transformed metrics relate to the geodesics of the original metric by means of a difference 1-form that involves only the differential of the conformal factor.

4. There are numerous analogies between the geodesics of optics and the geodesics of mechanics that are generally based in the assumption that the dynamical metric is conformally related to the kinematical one in various ways.

5. When a space-time is static and spatially-isotropic, it can be put into a form that is conformally equivalent to:

$$(c_0\, d\tau)^2 = c_0^2\, (dt)^2 - n^2\, \delta_{ij}\, \theta^i \theta^j \tag{6.2}$$

whose geodesics can, in fact, include light rays.

6. The astronomical phenomenon of gravitational lensing near strongly-gravitating stellar bodies can be accounted for by introducing an effective index of refraction that includes the effects of gravity on the space-time metric, and thus on null geodesics.

7. The relative motion of an optical medium will change the metric on the ambient space into an effective "optical metric" of the Gordon type that has an associated index of refraction that differs from its rest value.

## References (*)

1. E. W. Marchand, *Gradient-index Optics*, Academic Press, NY, 1978.
2. A. S. Eddington, *Space, time, and gravitation*, Cambridge Univ. Press, 1920.
3*. T. Levi-Civita, "La teoria di Einstein e il principio di Fermat," Nuovo Cim. **16** (1918), 105-114.
4. J. Plebanski, "Electromagnetic waves in gravitational fields," Phys. Rev. **118** (1950), 1396-1408.
5. A. M. Volkov, A. A. Izmest'ev, and G. V. Skrotskii, "The propagation of electromagnetic waves in a Riemannian space," Soviet Physics JETP **32** (1971), 686-689.
6. F. de Felice, "On the gravitational field acting as an optical medium." Gen. Rel. Grav. **2** (1971), 347-357.
7. X.-J. Wu and C.-M. Xu, "Null geodesic equations equivalent to the geometrical optics equation," Comm. in Theor. Phys. (Beijing) **9** (1988), 119-125.
8. V. Perlick, *Ray Optics, Fermat's Principle, and Applications to General Relativity*, Springer, Berlin, 2000.
9. X.-H. Ye, Q. Lin, "Gravitational lensing analyzed by graded refractive index of vacuum," J. Opt. A: Pure Appl. Optics **10** (2008) 075001.
10*. W. Gordon, "Zur Lichtfortpflanzung nach der Relativitätstheorie," Ann. Phys. (Leipzig) **72** (1923), 421-456.
11. M. v. Laue, *Die Relativitätstheorie*, v. I, 4th ed., Vieweg and Son, Braunschweig, 1921.
12. F. W. Hehl and Y. N. Obukhov, *Foundations of Classical Electrodynamics,* Birkhäuser, Boston, 2003.
13. D. H. Delphenich, *Pre-metric Electromagnetism*, Neo-classical Press, Spring Valley, OH, 2009.
14. T. Frenkel, *The Geometry of Physics: an introduction*, Cambridge University Press, Cambridge, 1997.
15. F. Warner, *Differentiable Manifolds and Lie Groups,* Scott Foresman, Glenview, IL, 1971.
16. M. Herzberger, *Strahlenoptik*, Springer, Berlin, 1931.
17. M. Kline and I. W. Kay, *Electromagnetic Theory and Geometrical Optics*, Interscience, NY, 1965.
18. M. Born and E. Wolf, *Principles of Optics,* Pergamon, Oxford, 1980.
19. C. G. J. Jacobi, *Vorlesungen über Dynamik*, Georg Reimer, Berlin, 1866, Lecture Six.

---

(*) References marked with an asterisk are available in English translation at the author's website.

20[*]. T. Levi-Civita, "Statica einsteiniana," Rend. Reale Accad. Lincei (5) **26** (1917), 458-470.
21. M. Abraham:
    – "Sulla teoria della gravitazione," Rend. Reale Accad. dei Lincei **20** (2[nd] sem., 1911), pp. 679-682.
    – "Una nuova theoria della gravitazione," Nuovo Cim. **4** (1912), pp. 459-481.
    – "Le equazioni di Lagrange nella nuova meccanica," Ann. di Mat. pura ed appl., Lagrange Centenary volume (3) **20** (1913), pp. 29-35.
22[*]. B. Caldonazzo, "Traiettorie dei raggi luminosi e dei punti materiali nel campo gravitazionale," Nuovo Cim. (5) **5** (1913), 267-300.
23. E. Schrödinger, "Hertz'sche Mechanik und Einstein'sche Gravitationstheorie," unpublished notes from around 1918.
24. J. Evans and M. Rosenquist, "F = ma optics," Am. J. Phys. 54 (1986), 876-883.
25[*]. A. Lichnerowicz, *Théories relativistes de la gravitation et de l'electromagnetisme,* Masson and Co., Paris, 1955.
26. J. Evans, K. K. Nandi, and A. Islam, "The optical-mechanical analogy in general relativity: exact Newtonian forms for the equations of motion of particles and photons," Gen. Rel. Grav. **28** (1996), 413-439.

__________